\documentclass{article}
\usepackage{spconf,amsmath,graphicx}

\usepackage{multirow}
\usepackage{booktabs}
\usepackage[title]{appendix}
\usepackage{url}
\usepackage{color}
\usepackage{dsfont}
\usepackage{amssymb}
\usepackage{caption}
\definecolor{grey}{rgb}{0.5, 0.5, 0.5}

\newcommand{\omitme}[1]{}

\title{TOWARDS LEARNING UNIVERSAL AUDIO REPRESENTATIONS}
\name{\parbox{0.95\linewidth}{\centering Luyu Wang, Pauline Luc, Yan Wu, Adri\`a Recasens, Lucas Smaira, Andrew Brock, Andrew Jaegle, \\ Jean-Baptiste Alayrac, Sander Dieleman, Joao Carreira, A\"aron van den Oord}}
\address{DeepMind, London, UK}
\begin{document}
%
\maketitle
\begin{abstract}
The ability to learn universal audio representations that can solve diverse speech, music, and environment tasks can spur many applications that require general sound content understanding.
In this work, we introduce a holistic audio representation evaluation suite (HARES) spanning 12 downstream tasks across audio domains and provide a thorough empirical study of recent sound representation learning systems on that benchmark.
We discover that previous sound event classification or speech models do not generalize outside of their domains.
We observe that more robust audio representations can be learned with the SimCLR objective; however, the model's transferability depends heavily on the model architecture.
We find the Slowfast architecture is good at learning rich representations required by different domains, but its performance is affected by the normalization scheme.
Based on these findings, we propose a novel normalizer-free Slowfast NFNet and achieve state-of-the-art performance across all domains.
\end{abstract}
\section{Introduction}
\label{sec:intro}



Audio is a ubiquitous form of data that can convey rich information, whether in the form of speech, music, or surrounding environment sounds.
An ability to understand all forms of sound in a holistic manner would empower a wide range of diverse applications, from end-to-end audio content understanding to human-computer interactions.
Importantly, when used on a new task, such an ability could reduce the amount of labeled data required, thanks to similarities with other domains and synergies with tasks where data is abundant.
In this paper, we take a step in this direction by laying the groundwork for learning universal audio representations.

We first turn our attention to evaluation benchmarks, and seek a compact and representative way to measure performance of universal audio representations.
In the text domain, GLUE \cite{wang2018glue} has been effective in driving the progress of unified approaches, and we follow a similar approach here.
Meta-benchmarks, such as NOSS \cite{shor2020towards} and SUPERB \cite{yang2021superb} have been introduced with similar motivation, but are restricted to the speech domain.
The AudioSet benchmark \cite{gemmeke2017audio} has greatly accelerated progress in sound understanding thanks to its scale and difficulty; but similarly, it provides an incomplete picture on the usefulness of representations.
Likewise, MagnaTagATune \cite{law2009evaluation} and NSynth \cite{pmlr-v70-engel17a} are limited to the music domain.
These benchmarks have facilitated model development in each individual domain \cite{jansen2018unsupervised, kong2019panns, schneider2019wav2vec, baevski2020wav2vec, wang2020contrastive,  gong2021ast, wang2021multi}.
A natural question is whether optimizing for performance on these different domains jointly can provide benefits in each domain.
The task suites used in \cite{tagliasacchi2020pre, saeed2020contrastive, niizumi2021byol} constitute early steps in this direction.
We build on their proposed benchmark, removing tasks where we noted performance saturation, and including additional challenging tasks, such as AudioSet \cite{gemmeke2017audio}, MagnaTagATune \cite{law2009evaluation}, and Fluent Speech Commands \cite{lugosch2019speech}.

Equipped with this benchmark, we perform thorough empirical studies on recent frameworks and architectures.
We first perform an extensive empirical comparison of training objectives and models, especially focusing on the two dominant paradigms of supervised classification \cite{kong2019panns, gong2021ast} and self-supervised contrastive learning \cite{saeed2020contrastive, wang2021multi}, and comparing them across a large set of model architectures.
We find that models trained with contrastive learning tend to generalize better in the speech and music domain, while performing comparably to supervised pretraining for environment sounds.
We also measure the performance of heavily-tuned speech representation systems \cite{oord2018representation, schneider2019wav2vec, baevski2020wav2vec}, but we find that their strong performance does not extend to other domains.
Motivated by the complementary strengths of Slowfast ResNet50 \cite{kazakos2021slow} and NFNet-F0 \cite{brock2021high}, we combine their design principles and evaluate a novel architecture, Slowfast NFNet-F0, which obtains improved performance across the board.
We additionally provide a thorough study on how normalization schemes affect the representations.
Putting these all together, we obtain better scores across the speech, music, and environment domain, as measured by our meta benchmark.

To sum up, we make the following contributions: (i) We introduce HARES, an evaluation benchmark designed to drive progress for universal audio representations; (ii) We perform empirical studies on models and learning frameworks. (iii) Based on our findings, we propose a novel architecture, Slowfast NFNet-F0, which significantly improves performance across the board.
(iv) Our final model obtains state-of-the-art results, in particular improving by a large margin in the music domain.

\section{Method}
\label{sec:method}

\subsection{Holistic Audio Representation Evaluation Suite}

\begin{table}[t]
    \centering
    \caption{\textbf{HARES tasks}. HARES unifies existing datasets with the goal of covering the three major audio domains to facilitate the development of models for general auditory content understanding.}
    \resizebox{0.48\textwidth}{!}{
    \begin{tabular}{lcccc}
    \toprule
        \textbf{Dataset} & \textbf{Task} & \textbf{\#Samples} & \textbf{\#Classes} & \textbf{Used} in \\
        \midrule
        \emph{Environment} & & \\
        AudioSet & Audio tagging & 1.9m & 527 & \cite{jansen2018unsupervised, kong2019panns, gong2021ast} \\
        Birdsong & Animal sound & 36k & 2 & \cite{tagliasacchi2020pre, saeed2020contrastive} \\
        TUT18 & Acoustic Scenes & 8.6k & 10 & \cite{tagliasacchi2020pre, saeed2020contrastive} \\
        ESC-50 & Acoustic Scenes & 2.0k & 50 & \cite{kong2019panns, wang2021multi} \\
        \midrule
        \emph{Speech} & & \\
        Speech Comm. v1 & Keyword & 90k & 12 & \cite{shor2020towards, saeed2020contrastive, yang2021superb, niizumi2021byol} \\
        Speech Comm. v2 & Keyword & 96k & 35 & \cite{tagliasacchi2020pre, saeed2020contrastive, niizumi2021byol} \\
        Fluent Speech Comm. & Intention & 27k & 31 & \cite{yang2021superb} \\
        VoxForge & Language id & 145k & 6 & \cite{shor2020towards, saeed2020contrastive, niizumi2021byol} \\
        VoxCeleb & Speaker id & 147k & 1251 & \cite{shor2020towards, saeed2020contrastive, yang2021superb, niizumi2021byol} \\
        \midrule
        \emph{Music} & & \\
        NSynth Instrument & Instrument id & 293k & 11 & \cite{tagliasacchi2020pre, saeed2020contrastive, niizumi2021byol} \\
        NSynth Pitch & Pitch estimation & 293k & 128 & - \\
        MagnaTagATune & Music tagging & 26k & 50 & \cite{spijkervet2021contrastive} \\
        \bottomrule
    \end{tabular}
    }
    \label{tab:hares}
\end{table}

Intuitively, sound can be categorized into three domains: speech, music, and environment (sound of things, animal, nature, etc).
Previous evaluation benchmarks either contain saturated tasks and lack coverage in the music and environment domain \cite{tagliasacchi2020pre, saeed2020contrastive}, or focus entirely on the speech domain \cite{shor2020towards, yang2021superb}.
We introduce a new evaluation meta-benchmark, named the Holistic Audio Representation Evaluation Suite (HARES) that contains 12 tasks spanning all three domains.\footnote{More details about HARES tasks and the Slowfast NFNet-F0 model can be found at \url{https://github.com/deepmind/slowfast_nfnets}.}
We build on top of tasks previously used in \cite{tagliasacchi2020pre, saeed2020contrastive}.
We exclude tasks where performance is saturated (LibriSpeech-speaker-id and MUSAN), and add widely used and challenging tasks, including AudioSet \cite{gemmeke2017audio}, ESC-50 \cite{piczak2015esc}, Fluent Speech Commands \cite{lugosch2019speech}, NSynth-pitch \cite{pmlr-v70-engel17a}, and MagnaTagATune \cite{law2009evaluation}.
We summarize all task characteristics (description, relevant statistics and occurrences in related work) in Table~\ref{tab:hares}.
Each of these benchmarks has been effective in driving model development, and we propose to jointly optimize performance across domains to obtain universal representations.
HARES shares 4 tasks with the SUPERB speech benchmark \cite{yang2021superb}.
We exclude speech and phoneme recognition tasks because the labels are temporally structured and they require sequence-to-sequence modeling. The models need to output representations with a high temporal resolution and it restricts the types of model that can be evaluated.

We evaluate by training a \emph{linear} layer on the \emph{frozen} pretrained model for all tasks, except for AudioSet where we follow the common 1-hidden-layer MLP protocol \cite{jansen2018unsupervised, wang2021multi, wang2021multimodal}.
Global average pooling is applied to the representations from the last layer before the fine-tune module.
We report the mean average precision (mAP) on AudioSet and MagnaTagATune due to their multi-class nature, and accuracy on all other datasets.
Note that although there is an overlap in the tasks, our results on speech tasks are not directly comparable to the SUPERB benchmark \cite{yang2021superb}, where the authors use weighted-sum of hidden representations from different layers and an additional linear projection layer for the downstream evaluation.
Our choice follows \cite{tagliasacchi2020pre, saeed2020contrastive} and keeps homogeneity with the rest.

\subsection{Training Frameworks and Architectures}

We perform an extensive empirical comparison of training objectives and models. We use AudioSet \cite{gemmeke2017audio} as the training dataset, due to its combined advantages of size and availability of labels for the supervised setting.

First, we focus our study on two dominant paradigms for pretraining: supervised classification \cite{kong2019panns, gong2021ast} and self-supervised contrastive learning (SimCLR) \cite{saeed2020contrastive, wang2021multi, wang2021multimodal, spijkervet2021contrastive}.
We compare the two training objectives across a large set of models used recently for audio representation learning, listed in Table~\ref{tab:arch_results}.
Note that most models are adopted from the vision domain; they treat spectrograms as images and do not distinguish time and frequency dimensions.
We use example mixing to augment spectrograms, whose effectiveness has been demonstrated in both settings \cite{kong2019panns, wang2021multi}.
Also, for sound classification, techniques including class-balanced data loading and model ensembles can boost the performance on the AudioSet benchmark \cite{kong2019panns, gong2021ast}, and pretraining on large image datasets can also significantly improve the performance of methods like ViT \cite{gong2021ast}.
For the sake of simplicity, we leave the exploration of such orthogonal directions for future work.

Second, we extend our study to other strongly performing approaches that explicitly model the time dimension, namely BYOL-A \cite{niizumi2021byol}, CPC \cite{oord2018representation, schneider2019wav2vec, wang2020contrastive} and masked prediction using Wav2Vec2.0 \cite{baevski2020wav2vec}.
We reproduce these methods to evaluate them on our benchmark.

\subsection{Slowfast NFNets}
\label{sec:slowfast-NFNets}

Due to its strong performance, we build on the Slowfast network architecture \cite{kazakos2021slow}, and seek to further improve its performance in the audio domain.
It uses separable convolutions to model the time and frequency dimension independently, and two pathways to capture both slow and fast features in the spectrogram.
Due to space limitations, we refer the reader to \cite{kazakos2021slow} for details.

\begin{table*}[t]
    \centering
    \caption{\textbf{Evaluating frameworks and architectures on HARES.} We compare the impact of architecture choice under the classification and SimCLR objective. We also show the performance of several other recent strongly performing frameworks. Average scores are reported for tasks in each domain separately, and all three combined. All models are trained on AudioSet except for bidirectional CPC and Wav2Vec2.0, for which we also show results when they are trained on LibriSpeech (LS).
    }
    \resizebox{0.82\textwidth}{!}{
    \begin{tabular}{lccc|ccccc}
    \toprule
        \textbf{Architecture} & \textbf{\#Params} & \textbf{Input format} & \textbf{Used in} & \textbf{Env.} & \textbf{Speech} & \textbf{Music} & \textbf{HARES} & \textbf{AudioSet (mAP)} \\
        \midrule
         & & & & \multicolumn{2}{c}{\emph{Classification/SimCLR}} & &  \\
        BYOL-A CNN  & 5.3m & Spectrogram & \cite{niizumi2021byol} & $69.8/69.9$ & $61.4/69.8$ & $57.6/63.1$ & $63.3/68.2$ & $33.8/32.2$ \\
        EfficientNet-B0 & 4.0m & Spectrogram & \cite{saeed2020contrastive}  & $70.5/63.8$ & $43.5/40.7$ & $48.0/44.0$ & $53.6/49.2$ & $32.1/26.2$ \\
        PANNs-CNN14 & 71m & Spectrogram & \cite{kong2019panns, wang2021multi} & $74.8/66.4$ & $56.0/37.3$ & $56.4/44.8$ & $62.4/48.9$ & $38.9/28.8$ \\
        ViT-Base & 86m & Spectrogram & \cite{gong2021ast} & $73.5/74.6$ & $50.4/56.5$ & $60.3/64.2$ & $60.6/64.5$ & $37.6/36.8$ \\
        ResNet50 & 23m & Spectrogram & \cite{jansen2018unsupervised} & $75.1/74.4$ & $51.7/65.0$ & $59.6/63.7$ & $61.4/67.8$ & $\underline{39.4}/36.2$ \\
        SF ResNet50 & 26m & Spectrogram & \cite{kazakos2021slow} & $74.3/74.3$ & $56.9/73.4$ & $59.6/65.2$ & $63.4/\underline{71.7}$ & $38.2/36.6$ \\
        NFNet-F0 & 68m & Spectrogram & Ours & $\mathbf{76.2}/\underline{76.0}$ & $59.0/65.9$ & $61.8/\underline{65.5}$ & $65.5/69.2$ & $\mathbf{39.6}/37.6$ \\
        SF NFNet-F0 & 63m & Spectrogram & Ours & $75.4/75.8$ & $65.6/\mathbf{77.2}$ & $64.5/\mathbf{68.6}$ & $68.6/\mathbf{74.6}$ & $38.7/37.8$ \\
        \midrule
        Bidir-CPC & 38m & Waveform & \cite{wang2020contrastive}    & $69.1$ & $72.4$ & $57.6$ & $67.6$ & $28.7$ \\
        Bidir-CPC (LS) & 38m & Waveform & \cite{wang2020contrastive}    & $59.9$ & $75.3$ & $51.5$ & $64.2$ & $21.3$ \\
        Wav2Vec2.0 & 95m & Waveform & \cite{baevski2020wav2vec}  & $48.3$ & $41.2$ & $35.4$ & $42.1$ & $14.8$ \\
        Wav2Vec2.0 (LS) & 95m & Waveform & \cite{baevski2020wav2vec}  & $48.5$ & $\underline{75.6}$ & $35.5$ & $56.6$ & $14.8$ \\
        BYOL-A & 5.3m & Spectrogram & \cite{niizumi2021byol} & $65.4$ & $66.9$ & $60.7$ & $64.9$ & $26.1$ \\
        \bottomrule
    \end{tabular}
    }
    \label{tab:arch_results}
\end{table*}

From our study on various architectures, we notice that the NFNet architecture \cite{brock2021high}, which is normalizer-free, performs very well. Motivated by this finding, we perform a thorough study on various normalization schemes for Slowfast networks. We compare Batch Norm \cite{ioffe2015batch}, Layer Norm \cite{ba2016layer}, Instance Norm \cite{ulyanov2016instance} as well as not using any normalizer. Finally, we combine the design principles of NFNet with those of the Slowfast network, to propose a novel architecture, denoted Slowfast NFNet-F0. It is free from the side effects of normalizers \cite{brock2021characterizing} and is optimized for training speed on recent hardware accelerators \cite{brock2021high}.


\section{EMPIRICAL STUDY}
\label{sec:study}
We perform thorough empirical studies using the proposed HARES benchmark.
Unless further noted, we train classification models with batch size 1024 for 100k steps, as these models converge fast and can overfit easily.
We train SimCLR models with batch size 4096 for 200k steps.
All models are trained on AudioSet~\cite{gemmeke2017audio} with Adam \cite{kingma2014adam} and an initial learning rate of $2\times10^{-4}$ on a Cloud TPU v3 Pod slice with 32 cores.
For classification and SimCLR, we randomly crop 3-second windows for training.
We extract log-mel spectrograms with 80 features using a window size of 25 ms and a stride of 10 ms from a waveform sampled at 16kHz.
The projector for the SimCLR loss is a MLP with 3 hidden layers of size 4096.

\subsection{Architectures and Losses}

We report the results of our comparison between supervised classification and SimCLR loss in Table~\ref{tab:arch_results}.
In general, we observe that the widely used AudioSet benchmark \cite{gemmeke2017audio} does not correlate with our HARES benchmark - it aligns with the overall score on the environment sub-suite.
On the speech and music benchmarks SimCLR leads to better generalization across architectures, and they perform similarly on the environment benchmark, with the exception of EfficientNet-B0 and CNN14.
We think this is because labels presented in supervised pretraining bias the network towards extracting more overall and slow features than local traits.
Architectures from the vision domain show strong performance on environment tasks but get relatively low scores on speech tasks.
This is possibly because they treat the time-frequency spectrograms as images with two equal spatial dimensions, which may ignore local features present in the audio.
This hypothesis is supported by the fact that approaches that handle the two dimensions differently tend to perform better on speech tasks: Slowfast networks thank to the use of separable convolutions, and BYOL-A CNN thanks to its fully connected layers only operating on the frequency and channel dimension.


\begin{table}[t]
    \centering
    \caption{\textbf{Impact of normalizer on Slowfast networks} trained with the SimCLR objective. We consider Batch Norm (BN), Layer Norm (LN), Instance Norm (IN), and Normalizer-Free (NF) architectures. Latencies are given as the time required for a full training step on TPU.
    }
    \resizebox{0.48\textwidth}{!}{
    \begin{tabular}{lccccc}
    \toprule
        \textbf{Architectures} & \textbf{Env.} & \textbf{Speech} & \textbf{Music} & \textbf{HARES} & \textbf{TPUv3 Train} \\
        \midrule
        SF ResNet50 (no BN) &$74.1$ & $74.1$ & $65.6$ & $71.9$ & $131.1$ ms \\
        SF ResNet50 &$74.3$ & $73.4$ & $65.2$ & $71.7$ & $188.0$ ms \\
        SF ResNet50-LN &$74.0$ & $76.0$ & $\underline{67.1}$ & $73.1$ & $203.7$ ms \\
        SF ResNet50-IN &$73.7$ & $\mathbf{77.7}$ & $66.0$ & $\underline{73.5}$ & $207.9$ ms \\
        SF NF-ResNet50 &$\underline{74.0}$ & $75.7$ & $66.6$ & $72.9$ & $158.0$ ms \\
        SF NFNet-F0 &$\mathbf{75.8}$ & $\underline{77.2}$ & $\mathbf{68.6}$ & $\mathbf{74.6}$ & $280.9$ ms \\
        \bottomrule
    \end{tabular}
    }
    \label{tab:slowfast_results}
\end{table}

\begin{table*}[h]
  \caption{\textbf{Comparisons to the state-of-the-art on HARES}. Both Slowfast networks are trained with the SimCLR loss. Wav2Vec2.0 is trained on LibriSpeech (LS) because it does not perform well when trained on AudioSet. CLMR is trained on MagnaTagATune and Million Songs. All other models are trained on AudioSet. *No class-balanced dataloader.
  }
  \label{tab:sota}
  \centering
  \resizebox{\textwidth}{!}{
  \begin{tabular}{ l c c c c c c | c c c c}
    \toprule
    \multirow{2}{*}{\textbf{Tasks}} & \textbf{SF NFNet-F0} & \textbf{SF ResNet50} & \textbf{Wav2Vec2.0} & \textbf{Bidir-CPC} & \textbf{BYOL-A} & \textbf{PANNs-CNN14} & \textbf{COLA} & \textbf{BYOL-A} & \textbf{PANNs-CNN14} & \textbf{CLMR} \\
     & (ours) & \cite{kazakos2021slow} (ours) & \cite{baevski2020wav2vec} (ours, LS) & \cite{wang2020contrastive} (ours) & \cite{niizumi2021byol} (ours) & \cite{kong2019panns} (ours) &    \cite{saeed2020contrastive} & \cite{niizumi2021byol} & \cite{kong2019panns} & \cite{spijkervet2021contrastive} \\
    \midrule
    \emph{Environment} & & & & & & & & & \\
    AudioSet & $37.8$ & $36.6$ & $14.8$ & $28.7$ & $26.1$ & $\mathbf{38.9}$ & - & - & $37.5^*$ & -~~~\\
    Birdsong  & $77.6$ & $77.9$ & $60.9$ & $77.6$ & $77.1$ & $\mathbf{78.0}$ & $77.0$ & - & - & -~~~\\
    TUT18 & $\mathbf{96.8}$ & $92.3$ & $49.3$ & $90.9$ & $76.3$ & $88.9$ & $94.0$ & - & - & -~~~\\
    ESC-50 & $91.1$ & $90.5$ & $69.2$ & $79.2$ & $82.3$ & $\mathbf{93.6}$ & - & - & $90.8$ & -~~~\\
    Avg. & $\mathbf{75.8}$ & $74.3$ & $48.5$ & $69.1$ & $65.4$ & $74.8$ & - & - & - & -~~~\\
    \midrule
    \emph{Speech} & & & & & & & & & \\
    Speech Comm. v1 & $91.7$ & $88.8$ & $\mathbf{95.1}$ & $92.0$ & $90.1$ & $71.4$ & $71.7$ &  $91.0$ & - & -~~~\\
    Speech Comm. v2 & $93.0$ & $91.5$ & $\mathbf{94.8}$ & $92.4$ & $89.7$ & $76.4$ & $62.4$ &  $92.2$ & - & -~~~\\
    VoxForge & $90.4$ & $89.2$ & $\mathbf{91.2}$ & $89.1$ & $83.9$ & $78.5$ & $71.3$ &  $90.2$ & - & -~~~\\
    VoxCeleb & $\textbf{64.9}$ & $59.4$ & $41.6$ & $40.7$ & $39.5$ & $30.9$ & $29.9$ &  $40.1$ & - & -~~~\\
    Fluent Comm. & $46.1$ & $38.3$ & $\mathbf{55.3}$ & $47.9$ & $31.3$ & $22.8$ & - & - & - & -~~~\\
    Avg. & $\mathbf{77.2}$ & $73.4$ & $75.6$ & $72.4$ & $66.9$ & $56.0$ & - & - & - & -~~~\\
    \midrule
    \emph{Music} & & & & & & & & & \\
    NSynth-Pitch & $\mathbf{88.0}$ & $83.2$ & $34.6$ & $78.1$ & $71.4$ & $58.4$ & - & - & - & -~~~\\
    NSynth-Instrument & $\mathbf{78.2}$ & $73.5$ & $40.2$ & $58.4$ & $73.6$ & $71.4$ & $63.4$ &  $74.1$ & - & -~~~\\
    MagnaTagATune & $\mathbf{39.5}$ & $38.8$ & $31.7$ & $36.3$ & $37.1$ & $39.4$ & - & - & - & $36.0$~~~\\
    Avg. & $\mathbf{68.6}$ & $65.2$ & $35.5$ & $57.6$ & $60.7$ & $56.4$ & - & - & - & -~~~\\
    \midrule
    HARES & $\mathbf{74.6}$ & $71.7$ & $56.6$ & $67.6$ & $64.9$ & $62.4$ & - & - & - & -~~~\\
    \bottomrule
  \end{tabular}
  }
  
\end{table*}

Next, we replicate and evaluate the bidirectional version of CPC/Wav2Vec \cite{wang2020contrastive}, Wav2Vec2.0 \cite{baevski2020wav2vec}, and BYOL-A \cite{niizumi2021byol}.
All three models show competitive performance on speech tasks, but they do not transfer well on the other two domains.
CPC and Wav2Vec2.0 directly operate on raw audio sequences and use the contrastive objective over time steps: this procedure preserves the temporal structure of the data by design.
Interestingly, we find Wav2Vec2.0 does not perform well when trained with AudioSet.
This is possibly because the complexity of the data is beyond the capability of the masked prediction objective, which works the best on clean speech datasets such as LibriSpeech \cite{panayotov2015librispeech}.
BYOL-A \cite{niizumi2021byol} uses a Siamese framework but takes only one temporal crop at random and augments it with random resize cropping: this effectively biases the model towards local semantics.
Notably, we find that with the same CNN architecture, our SimCLR-based framework obtains much better performance than the original BYOL framework proposed in \cite{niizumi2021byol}.
We note that this could be in part due to the more extensive study we perform under the former setting.
An in-depth comparison between SimCLR and BYOL for audio is beyond the scope of this work.

Overall, NFNet-F0 performs remarkably on the environment sub-suite.
When comparing the Slowfast networks with their vanilla counterparts, we find that they are comparable on environment tasks, but better on the speech and music tasks. This shows the effectiveness of the Slowfast architecture.

\subsection{Normalizers Matter For Slowfast Networks}

We find that the choice of normalizer affects transferability.
The original Slowfast ResNet50 \cite{kazakos2021slow} for audio makes use of Batch Norm.
In Table~\ref{tab:slowfast_results}, we observe that using Layer Norm or Instance Norm, which do not break the independence between training examples in the batch, improves the scores in all domains.
We adapt the NF ResNet50 \cite{brock2021characterizing} and NFNet-F0 \cite{brock2021high} to the Slowfast setting.
The latter is better optimized for performance on recent hardware accelerators.
Even in the absence of any normalizer, Slowfast NF ResNet50 performs similarly to when layer or instance normalization is used instead.
Furthermore, our Slowfast NFNet-F0  outperforms all other models.

\subsection{Comparison to the State-of-the-Art}

We compare our best models to previous works in Table~\ref{tab:sota}.
Our Slowfast NFNet-F0 trained with the SimCLR objective outperforms all previous frameworks on the overall HARES benchmark as well as in each domain, including COLA \cite{saeed2020contrastive} and BYOL-A \cite{niizumi2021byol} previously developed for general-purpose audio understanding.
It outperforms the previously proposed Slowfast ResNet50 \cite{kazakos2021slow} on almost all tasks.
In addition, it also outperforms strongly performing frameworks developed in different audio domains.
It has a slight edge over the supervised PANNs-CNN14 sound classification model on environment tasks.
It performs slightly worse on semantic speech tasks than Wav2Vec2.0 (trained on LibriSpeech), but there is a substantial improvement on the VoxCeleb speaker identification task.
Meanwhile, it shows strong performance on the music sub-suite.
It achieves an audio-only state-or-the-art on the MagnaTagATune task with an mAP of 39.5, outperforming CLMR trained on a music corpus at 36.0 \cite{spijkervet2021contrastive}.




\section{Conclusion}
\label{sec:conclusion}


In this paper, we devised a task suite called HARES for holistic evaluation of audio representations.
We study several recent pre-training frameworks, focusing in particular on supervised classification and self-supervised contrastive learning with many different models.
We find that models trained with contrastive learning and Slowfast networks tend to generalize better across different domains.
We also perform a thorough study of different normalization schemes, building on the Slowfast network architecture, and obtain significant performance improvements using a new Slowfast NFNet-F0.
Our final model significantly outperforms previous general-purpose audio representation models.
We hope the HARES benchmark and the findings presented in this paper can spur future research towards further improving universal audio representations.




\bibliographystyle{IEEEbib}
\bibliography{refs}

\begin{appendices}

\section{HARES Tasks Details}
In this section, we provide more relevant details about HARES tasks in addition to the information shown in Table~\ref{tab:hares}. All of the HARES tasks are public available\footnote{\url{https://github.com/deepmind/slowfast_nfnets}}. We evaluate by training a \emph{linear} layer on the \emph{frozen} pretrained model for all tasks except for AudioSet, where we follow the common 1-hidden-layer MLP protocol on \emph{frozen} features \cite{jansen2018unsupervised}.
Global average pooling is applied before the fine-tune module. If the dataset has a validation set, we ignore it and only report the score on the test set. Unless further noted, all downstream jobs are trained with a total batch size of 64 for 400k steps on Cloud TPU v2 slice with 8 cores. We use the Adam optimizer~\cite{kingma2014adam}, first linearly increasing the learning rate from $0$ to $2\times10^{-4}$ over 5k steps, and then decaying it following a cosine schedule \cite{loshchilov2016sgdr} down to $0$. Except for AudioSet and ESC-50, we do not further apply augmentations to downstream evaluations.

\noindent \textbf{AudioSet} \cite{gemmeke2017audio}: Due the attrition of YouTube URLs\footnote{\url{research.google.com/audioset/}}, our version of AudioSet has around 1.9 million samples. Following \cite{jansen2018unsupervised}, the downsteam MLP has one hidden layer with 512 hidden units. We train and evaluate models using crops of 3 seconds. Same as previous works \cite{kong2019panns, wang2021multi, wang2021multimodal}, we apply example mixing to audio samples during training. During evaluation, we average the scores of 10 overlapped and equally spaced 3-second clips. AudioSet downstream tasks are trained with a batch size of 128 for 200k steps. For supervised classification, we do not perform a second stage of fine-tuning to evaluate the representations on the AudioSet benchmark.

\noindent \textbf{Birdsong} \cite{stowell2019automatic}: This dataset is constructed from three development datasets (freefield1010, warblrb10k, BirdVox-DCASE-20k) of the Task 3 of DCASE2018 challenge\footnote{\url{dcase.community/challenge2018/task-bird-audio-detection}}. Note that the training and test splits are different from ones provided by the challenge as labels of the original test set is not accessible. Instead, we use the splits created and provided by the authors of \cite{tagliasacchi2020pre}. Each sample is around 10 seconds long. We train models with 1-second crops. During evaluation, we average the scores over consecutive non-overlapped 1-second subclips that cover the whole clip.

\noindent \textbf{TUT18} \cite{heittola2018tut}: This dataset is created from the development dataset of Task 1 - Subtask A of DCASE2018 challenge\footnote{\url{dcase.community/challenge2018/task-acoustic-scene-classification}}. Similarly, the training and test splits are different from original ones. We use the splits created and provided by the authors of \cite{tagliasacchi2020pre}. Each sample is around 10 seconds long. We train models with 5-second crops. During evaluation, we average the scores over consecutive non-overlapped 5-second subclips covering the entire clip.

\noindent \textbf{ESC-50} \cite{piczak2015esc}: For the linear evaluation on ESC-50\footnote{\url{github.com/karolpiczak/ESC-50}} we follow the setup used in \cite{recasens2021broaden}. We use the SVM implementation of SciKit-Learn \cite{pedregosa2011scikit}. We use full 5-second clips for training and evaluation. For training, we process 10 epochs worth of augmented samples. We sweep the value for the SVM regularization parameter in the following set of values: $\{3\times10^{-5}, 10^{-4}, 3\times10^{-4}, 10^{-3}, 3\times10^{-3}, 10^{-2}, 3\times10^{-2}$. We use the first split to pick the optimal value and report the average of all the splits.

\noindent \textbf{Speech Commands} \cite{warden2018speech}: In this work we use both Speech Commands dataset V1\footnote{\url{download.tensorflow.org/data/speech_commands_v0.01.tar.gz}} and V2\footnote{\url{download.tensorflow.org/data/speech_commands_v0.02.tar.gz}} with 12 and 35 labels, respectively. We train and evaluate models with the entire 1-second audio clips.

\noindent \textbf{Fluent Speech Commands} \cite{lugosch2019speech}: In this dataset\footnote{\url{fluent.ai/research/fluent-speech-commands/}}, each audio is labeled with three slots. Unlike in \cite{lugosch2019speech}, we do not use any ASR module and directly use the model to predict the intent from the raw waveform. The model predicts three sets of logits upon which three cross entropy losses are computed simultaneously. The prediction is correct only if all three slots are predicted correctly. We train models with 3-second crops. During evaluation, we average the scores over consecutive non-overlapped 3-second subclips covering the whole audio.

\noindent \textbf{VoxForge} \cite{maclean2018voxforge}: This dataset is available through Tensorflow Dataset\footnote{\url{tensorflow.org/datasets/catalog/voxforge}}. We train models with 3-second crops. During evaluation, we average the scores over consecutive non-overlapped 3-second subclips that cover the whole clip.

\noindent \textbf{VoxCeleb} \cite{nagrani2017voxceleb}: This dataset is available through Tensorflow Dataset\footnote{\url{tensorflow.org/datasets/catalog/voxceleb}}. We train models with 3-second crops. During evaluation, we average the scores over consecutive non-overlapped 3-second subclips that cover the whole clip.

\noindent \textbf{NSynth} \cite{pmlr-v70-engel17a}: This dataset is available through Tensorflow Dataset\footnote{\url{tensorflow.org/datasets/catalog/nsynth}}. It is used for both the instrument classification and pitch estimation task. For instrument classification, we directly use the 4-second clip for both training and evaluation. For the pitch estimation task, we train models with 1-second crops, and during evaluation, we average the scores over four non-overlapped 1-second subclips.

\noindent \textbf{MagnaTagATune} \cite{law2009evaluation}: For this dataset\footnote{\url{mirg.city.ac.uk/codeapps/the-magnatagatune-dataset}}, we use the same splits\footnote{\url{github.com/jordipons/musicnn-training/tree/master/data/index/mtt}} as previous works \cite{pons2017end, spijkervet2021contrastive}. We train and evaluate models using crops of 3 seconds. During evaluation, we average the scores over consecutive non-overlapped 3-second subclips that cover the entire audio. We notice that the downstream training on this dataset converges quickly. Therefore, models are trained with a batch size of 128 for 20k steps.

\begin{table}[t]
    \centering
    \caption{\textbf{Slowfast NFNet-F0}. Slow and fast path has a group size of 128 and 16, respectively. Each block has a stride of $1, 2$ on the time and frequency dimension, respectively. Final features are obtained by global average pooling and concatenation of two pathways.}
    \resizebox{0.45\textwidth}{!}{
    \begin{tabular}{c|c|c|c}
    \toprule
        \textbf{Stage} & \textbf{Slow path} & \textbf{Fast path} & \textbf{$T\times F$} \\
        \toprule
        spectrogram & - & - & $400\times 128$ \\
        \midrule
        \multirow{2}{*}{data layer} & \multirow{2}{*}{stride $4,1$} & \multirow{2}{*}{stride $1,1$} & $Slow: 100\times 128$ \\
         & & & $Fast: 400\times 128$ \\
        \midrule
        \multirow{2}{*}{stem1} & $1\times3, 16$ & $3\times3, 2$ & $Slow: 50\times 64$ \\
         & stride $2,2$ & stride $2,2$ & $Fast: 200\times 64$ \\
        \midrule
        \multirow{2}{*}{stem2} & $1\times3, 32$ & $3\times3, 4$ & $Slow: 50\times 64$ \\
         & stride $1,1$ & stride $1,1$ & $Fast: 200\times 64$ \\
        \midrule
        \multirow{2}{*}{stem3} & $1\times3, 64$ & $3\times3, 8$ & $Slow: 50\times 64$ \\
         & stride $1,1$ & stride $1,1$ & $Fast: 200\times 64$ \\
        \midrule
        \multirow{2}{*}{stem4} & $3\times3, 128$ & $3\times3, 16$ & $Slow: 25\times 32$ \\
         & stride $2,2$ & stride $2,2$ & $Fast: 100\times 32$ \\
        \midrule
        block1 & $\begin{bmatrix}1\times1,128\\1\times1,128\\1\times3,128\\ 1\times1,256\end{bmatrix} \times 1 $ & $\begin{bmatrix}1\times1,16\\3\times1,16\\1\times3,16\\ 1\times1,32\end{bmatrix} \times 1 $ & $\begin{matrix}Slow: 25\times 32 \\ Fast: 100\times 32 \end{matrix}$ \\
        \midrule
        block2 & $\begin{bmatrix}1\times1,256\\1\times1,256\\1\times3,256\\ 1\times1,512\end{bmatrix} \times 2 $ & $\begin{bmatrix}1\times1,32\\3\times1,32\\1\times3,32\\ 1\times1,64\end{bmatrix} \times 2 $ & $\begin{matrix}Slow: 25\times 16 \\ Fast: 100\times 16 \end{matrix}$ \\
        \midrule
        block3 & $\begin{bmatrix}1\times1,768\\3\times1,768\\1\times3,768\\ 1\times1,1536\end{bmatrix} \times 6 $ & $\begin{bmatrix}1\times1,96\\3\times1,96\\1\times3,96\\ 1\times1,192\end{bmatrix} \times 6 $ & $\begin{matrix}Slow: 25\times 8 \\ Fast: 100\times 8 \end{matrix}$ \\
        \midrule
        block4 & $\begin{bmatrix}1\times1,768\\3\times1,768\\1\times3,768\\ 1\times1,1536\end{bmatrix} \times 3 $ & $\begin{bmatrix}1\times1,96\\3\times1,96\\1\times3,96\\ 1\times1,192\end{bmatrix} \times 3 $ & $\begin{matrix}Slow: 25\times 4 \\ Fast: 100\times 4 \end{matrix}$ \\
        \midrule
        \multicolumn{3}{c|}{Global average pool \& concatenate} & \emph{Feature dim}: 1728 \\
        \bottomrule
    \end{tabular}
    }
    \label{tab:sf-NFNet}
\end{table}

\section{Slowfast NFNet-F0 Architecture}
We combine the design principles of Slowfast networks \cite{kazakos2021slow} and NFNets \cite{brock2021high} to devise a Slowfast NFNet-F0\footnote{\url{https://github.com/deepmind/slowfast_nfnets/blob/main/slowfast_nfnet.py}}, whose architecture is shown in Table~\ref{tab:sf-NFNet}.
Following \cite{kazakos2021slow}, the slow stream, compared to the fast stream, has 8 times more channel capacity, whose input spectrogram is strided temporally by 4 and thus has a lower temporal resolution.
We also apply the multi-level fast-to-slow fusion before each block with the same configuration as \cite{kazakos2021slow}, except that the Batch Norm layer is removed, and Scaled Weight Standardization is applied to 2D convolutions \cite{qiao2019weight, brock2021characterizing}.
The ResNet blocks in the original Slowfast network \cite{kazakos2021slow} are replaced with NFNet blocks, and the stem modules are also changed to be the same as ones used in \cite{brock2021high}.
Each block has a stride of 1 and 2 on the time and frequency dimension, respectively.
We follow the original choices for block hyperparameters, except that the slow and fast path has a group size of 128 and 16 respectively, and the $3\times3$ convolutions are replaced by separable convolutions.
Final features are obtained by global average pooling and concatenation of two pathways.
We use Stochastic Depth \cite{huang2016deep} with a rate of 0.1.
Unlike the original NFNets we do not apply any Dropout \cite{srivastava2014dropout}.
Note that this design can be extended to adopting bigger versions of the NFNet architecture, which we leave for future work.

In our experiments, we do not experience instability when training NFNets with large batch sizes (maximum 4096). Therefore, we do not use the Adaptive Gradient Clipping techniques which is employed by the original NFNet paper for the vision domain \cite{brock2021high}.

\section{Additional Pre-Training Details}
For supervised and SimCLR, we use the Adam optimizer~\cite{kingma2014adam}, first linearly increasing the learning rate from $0$ to $2\times10^{-4}$ over 5k steps, and then decaying it following a cosine schedule \cite{loshchilov2016sgdr} down to $0$. All our SimCLR models are trained with a temperature $\tau$ of 0.1. The example mixing ratio is sampled on the fly from the $\beta(5, 2)$ distribution. For spectrogram-based models, we standardize each input spectrogram by its mean and standard deviation.

When pretraining Wav2Vec2.0 \cite{baevski2020wav2vec} and bi-directional CPC \cite{wang2020contrastive} models, we use crops of 10 seconds and a batch size of 256. We use the same optimizer setup as that used for SimCLR except that on LibriSpeech \cite{panayotov2015librispeech} we train the models for 80k steps. We do not use any audio augmentation for these two models. For Wav2Vec2.0, we use both continuous inputs and targets for the target network \cite{baevski2020wav2vec}. For other hyperparameters, we follow the same choices in the original papers \cite{baevski2020wav2vec, wang2020contrastive}. For BYOL-A, we follow the exact pre-training setup in the original paper \cite{niizumi2021byol}.

\section{Full experiment results}

Full results for Table~\ref{tab:arch_results} and Table~\ref{tab:slowfast_results} are shown in Table \ref{tab:cls}, \ref{tab:simclr}, \ref{tab:slowfast-cls}, \ref{tab:slowfast-simclr}.

\begin{table*}[h]
  \caption{\textbf{Performance of models trained with classification loss}.
  }
  \label{tab:cls}
  \centering
  \begin{tabular}{ l c c c c c c c}
    \toprule
    \textbf{Architecture} & \textbf{BYOL-A} & \textbf{EfficientNet-B0}   & \textbf{ResNet50} &  \textbf{CNN14} & \textbf{NFNet-F0} & \textbf{ViT-Base} \\
    \midrule
    \emph{Environment} & & \\
    AudioSet & $33.8$ & $32.1$ & $39.4$ & $38.9$ & $39.6$ &  $37.6$~~~\\
    Birdsong  & $77.0$ & $75.8$ & $77.7$ & $78.0$ & $78.4$ &  $76.6$~~~\\
    TUT18 & $83.3$ & $83.3$ & $89.6$ & $88.9$ & $92.3$ &  $87.4$~~~\\
    ESC-50 & $84.9$ & $90.9$ & $93.5$ & $93.6$ & $94.6$ &  $92.6$~~~\\
    Avg. & $69.8$ & $70.5$ & $75.1$ & $74.8$ & $76.2$ &  $73.5$~~~\\
    \midrule
    \emph{Speech} & & & \\
    Speech Comm. v1 & $81.8$ & $58.6$ & $64.4$ & $71.4$ & $75.2$ &  $62.4$~~~\\
    Speech Comm. v2 & $85.1$ & $60.1$ & $69.0$ & $76.4$ & $79.9$ &  $70.1$~~~\\
    VoxForge & $78.6$ & $70.3$ & $75.5$ & $78.5$ & $79.3$ &  $74.7$~~~\\
    VoxCeleb & $33.4$ & $13.8$ & $30.1$ & $30.9$ & $35.9$ &  $29.6$~~~\\
    Fluent Comm. & $28.3$ & $14.6$ & $19.3$ & $22.8$ & $24.9$ &  $15.0$~~~\\
    Avg. & $61.4$ & $43.5$ & $51.7$ & $56.0$ & $59.0$ &  $50.4$~~~\\
    \midrule
    \emph{Music} & & & \\
    NSynth-Pitch & $64.4$ & $41.5$ & $69.3$ & $58.4$ & $73.9$ &  $71.6$~~~\\
    NSynth-Instrument & $70.1$ & $64.4$ & $70.7$ & $71.4$ & $73.0$ &  $70.7$~~~\\
    MagnaTagATune & $38.4$ & $38.1$ & $38.7$ & $39.4$ & $38.6$ & $38.7$~~~\\
    Avg. & $57.6$ & $48.0$ & $59.6$ & $56.4$ & $61.8$ &  $60.3$~~~\\
    \midrule
    HARES & $63.3$ & $53.6$ & $61.4$ & $62.4$ & $65.5$ &  $60.6$~~~\\
    \bottomrule
  \end{tabular}
  
\end{table*}

\begin{table*}[h]
  \caption{\textbf{Performance of models trained with SimCLR loss}.
  }
  \label{tab:simclr}
  \centering
  \begin{tabular}{ l c c c c c c c}
    \toprule
    \textbf{Architecture} & \textbf{BYOL-A} & \textbf{EfficientNet-B0}   & \textbf{ResNet50} &  \textbf{CNN14} & \textbf{NFNet-F0} & \textbf{ViT-Base} \\
    \midrule
    \emph{Environment} & & \\
    AudioSet & $32.2$ & $26.2$ & $36.2$ & $28.8$ & $37.6$ &  $36.8$~~~\\
    Birdsong  & $76.6$ & $74.2$ & $77.7$ & $76.4$ & $77.5$ &  $77.1$~~~\\
    TUT18 & $83.8$ & $76.8$ & $94.0$ & $82.2$ & $96.9$ &  $93.7$~~~\\
    ESC-50 & $87.0$ & $77.9$ & $89.7$ & $78.0$ & $92.2$ &  $90.6$~~~\\
    Avg. & $69.9$ & $63.8$ & $74.4$ & $66.4$ & $76.0$ &  $74.6$~~~\\
    \midrule
    \emph{Speech} & & & \\
    Speech Comm. v1 & $89.5$ & $44.5$ & $78.9$ & $43.2$ & $82.3$ &  $65.1$~~~\\
    Speech Comm. v2 & $91.4$ & $55.4$ & $83.2$ & $44.7$ & $86.2$ &  $74.6$~~~\\
    VoxForge & $84.3$ & $67.5$ & $84.4$ & $68.2$ & $86.0$ &  $83.7$~~~\\
    VoxCeleb & $55.0$ & $22.1$ & $50.0$ & $19.1$ & $50.1$ &  $44.4$~~~\\
    Fluent Comm. & $28.9$ & $14.0$ & $28.3$ & $11.3$ & $24.8$ &  $14.8$~~~\\
    Avg. & $69.8$ & $40.7$ & $65.0$ & $37.3$ & $65.9$ &  $56.5$~~~\\
    \midrule
    \emph{Music} & & & \\
    NSynth-Pitch & $76.1$ & $48.0$ & $78.5$ & $36.1$ & $82.9$ &  $81.1$~~~\\
    NSynth-Instrument & $75.0$ & $58.0$ & $73.8$ & $63.7$ & $74.0$ &  $71.8$~~~\\
    MagnaTagATune & $38.2$ & $26.0$ & $38.7$ & $34.6$ & $39.6$ & $39.7$~~~\\
    Avg. & $63.1$ & $44.0$ & $63.7$ & $44.8$ & $65.5$ &  $64.2$~~~\\
    \midrule
    HARES & $68.2$ & $49.2$ & $67.8$ & $48.9$ & $69.2$ &  $64.5$~~~\\
    \bottomrule
  \end{tabular}
  
\end{table*}

\begin{table*}[h]
  \caption{\textbf{Performance of Slowfast models trained with classification loss}.
  }
  \label{tab:slowfast-cls}
  \centering
  \begin{tabular}{ l c c c c c c c}
    \toprule
    \multirow{2}{*}{\textbf{Architecture}} & \multicolumn{4}{c}{\textbf{SF ResNet50}} & \multirow{2}{*}{\textbf{SF NF-ResNet50}} & \multirow{2}{*}{\textbf{SF NFNet-F0}}  \\\cline{2-5}
     & \textbf{no norm} & \textbf{BN} & \textbf{LN} & \textbf{IN} &  &  \\
    \midrule
    \emph{Environment} & & \\
    AudioSet & $36.4$ & $38.2$ & $37.9$ & $34.8$ & $36.2$ &  $38.7$~~~\\
    Birdsong  & $77.3$ & $77.8$ & $77.7$ & $77.5$ & $77.2$ &  $77.4$~~~\\
    TUT18 & $90.7$ & $89.4$ & $91.9$ & $87.0$ & $91.2$ &  $92.7$~~~\\
    ESC-50 & $90.8$ & $91.8$ & $92.8$ & $90.4$ & $90.9$ &  $92.7$~~~\\
    Avg. & $73.8$ & $74.3$ & $75.1$ & $72.4$ & $73.9$ &  $75.4$~~~\\
    \midrule
    \emph{Speech} & & & \\
    Speech Comm. v1 & $86.9$ & $73.2$ & $78.4$ & $74.7$ & $83.2$ &  $83.0$~~~\\
    Speech Comm. v2 & $88.9$ & $78.2$ & $83.1$ & $80.1$ & $85.6$ &  $86.2$~~~\\
    VoxForge & $80.4$ & $76.6$ & $77.8$ & $69.5$ & $80.7$ &  $82.9$~~~\\
    VoxCeleb & $37.7$ & $31.9$ & $32.2$ & $22.6$ & $39.2$ &  $40.2$~~~\\
    Fluent Comm. & $37.1$ & $24.6$ & $26.5$ & $25.6$ & $35.0$ &  $35.5$~~~\\
    Avg. & $66.2$ & $56.9$ & $59.6$ & $54.5$ & $64.7$ &  $65.6$~~~\\
    \midrule
    \emph{Music} & & & \\
    NSynth-Pitch & $76.6$ & $71.1$ & $75.3$ & $67.2$ & $77.3$ &  $79.9$~~~\\
    NSynth-Instrument & $70.8$ & $69.6$ & $73.8$ & $71.2$ & $72.0$ &  $75.0$~~~\\
    MagnaTagATune & $38.4$ & $38.1$ & $38.7$ & $38.8$ & $38.0$ & $38.5$~~~\\
    Avg. & $61.9$ & $59.6$ & $62.6$ & $59.1$ & $62.4$ & $64.5$~~~\\
    \midrule
    HARES & $67.7$ & $63.4$ & $65.5$ & $61.6$ & $67.2$ & $68.6$~~~\\
    \bottomrule
  \end{tabular}
  
\end{table*}

\begin{table*}[h]
  \caption{\textbf{Performance of Slowfast models trained with SimCLR loss}.
  }
  \label{tab:slowfast-simclr}
  \centering
  \begin{tabular}{ l c c c c c c c}
    \toprule
    \multirow{2}{*}{\textbf{Architecture}} & \multicolumn{4}{c}{\textbf{SF ResNet50}} & \multirow{2}{*}{\textbf{SF NF-ResNet50}} & \multirow{2}{*}{\textbf{SF NFNet-F0}}  \\\cline{2-5}
     & \textbf{no norm} & \textbf{BN} & \textbf{LN} & \textbf{IN} &  &  \\
    \midrule
    \emph{Environment} & & \\
    AudioSet & $35.9$ & $36.6$ & $36.2$ & $36.0$ & $36.2$ &  $37.8$~~~\\
    Birdsong  & $77.9$ & $77.9$ & $78.0$ & $77.3$ & $78.0$ &  $77.6$~~~\\
    TUT18 & $91.9$ & $92.3$ & $92.0$ & $91.9$ & $93.0$ & $96.8$~~~\\
    ESC-50 & $90.6$ & $90.5$ & $90.0$ & $89.5$ & $88.7$ & $91.1$~~~\\
    Avg. & $74.1$ & $74.3$ & $74.0$ & $73.7$ & $74.0$ &  $75.8$~~~\\
    \midrule
    \emph{Speech} & & & \\
    Speech Comm. v1 & $89.7$ & $88.8$ & $91.1$ & $91.8$ & $91.0$ &  $91.7$~~~\\
    Speech Comm. v2 & $92.0$ & $91.5$ & $93.2$ & $93.9$ & $92.2$ &  $93.0$~~~\\
    VoxForge & $87.7$ & $89.2$ & $89.2$ & $86.6$ & $89.9$ & $90.4$~~~\\
    VoxCeleb & $59.5$ & $59.4$ & $60.6$ & $67.9$ & $61.1$ & $64.9$~~~\\
    Fluent Comm. & $41.5$ & $38.3$ & $45.9$ & $48.5$ & $44.3$ &  $46.1$~~~\\
    Avg. & $74.1$ & $73.4$ & $76.0$ & $77.7$ & $75.7$ &  $77.2$~~~\\
    \midrule
    \emph{Music} & & & \\
    NSynth-Pitch & $82.4$ & $83.2$ & $85.9$ & $84.5$ & $84.4$ &  $88.0$~~~\\
    NSynth-Instrument & $75.2$ & $73.5$ & $76.1$ & $74.3$ & $76.6$ &  $78.2$~~~\\
    MagnaTagATune & $39.1$ & $38.8$ & $39.2$ & $39.3$ & $38.9$ & $39.5$~~~\\
    Avg. & $65.6$ & $65.2$ & $67.1$ & $66.0$ & $66.6$ & $68.6$~~~\\
    \midrule
    HARES & $71.9$ & $71.7$ & $73.1$ & $73.5$ & $72.9$ & $74.6$~~~\\
    \bottomrule
  \end{tabular}
  
\end{table*}

\end{appendices}

\end{document}